\documentclass[conference]{IEEEtran}
\IEEEoverridecommandlockouts
\usepackage{cite}
\usepackage{amsmath,amssymb,amsfonts}
\usepackage{algorithmic}
\usepackage{graphicx}
\usepackage{float}
\usepackage{textcomp}
\usepackage{xcolor}
\usepackage{subfigure}
\usepackage{url}
\usepackage{algorithm}
\usepackage{algorithmic}
\usepackage{makecell}
\usepackage{color}
\usepackage{verbatim}
\usepackage{hyperref}
\usepackage{wrapfig}
\usepackage{multicol}
\usepackage{fancyhdr}

\def\BibTeX{{\rm B\kern-.05em{\sc i\kern-.025em b}\kern-.08em
    T\kern-.1667em\lower.7ex\hbox{E}\kern-.125emX}}

\begin{document}

\pagestyle{fancy}

\fancyhead[C]{This paper appears in IEEE Global Communications Conference (GLOBECOM) 2023.}
\title{Heterogeneous 360 Degree Videos in Metaverse: Differentiated Reinforcement Learning Approaches}

\author{
\IEEEauthorblockN{
Wenhan Yu,
Jun Zhao
}
\IEEEauthorblockA{
Nanyang Technological University \\
\IEEEauthorblockA{wenhan002@e.ntu.edu.sg, junzhao@ntu.edu.sg}
}
}

\maketitle
\thispagestyle{fancy}
\begin{abstract}
Advanced video technologies are driving the development of the futuristic Metaverse, which aims to connect users from anywhere and anytime. As such, the use cases for users will be much more diverse, leading to a mix of 360-degree videos with two types: \textit{non-VR} and \textit{VR} 360$^{\circ}$ videos. This paper presents a novel Quality of Service model for heterogeneous 360$^{\circ}$ videos with different requirements for frame rates and cybersickness. We propose a frame-slotted structure and conduct frame-wise optimization using self-designed differentiated deep reinforcement learning algorithms. Specifically, we design two structures, Separate Input Differentiated Output (SIDO) and Merged Input Differentiated Output (MIDO), for this heterogeneous scenario. We also conduct comprehensive experiments to demonstrate their effectiveness.

\end{abstract}

\begin{IEEEkeywords}
Metaverse, resource allocation, reinforcement learning, wireless networks.
\end{IEEEkeywords}

\section{Introduction}
\textbf{Background.} Virtual Reality (VR) is a crucial tool for creating a fully immersive and interactive experience for users, particularly in the context of the Metaverse that seeks to integrate all users into a unified and comprehensive virtual world. $360$-degree video (360$^{\circ}$ video) is a vital technology for VR, as it provides users with an immersive and intuitive way to explore virtual environments. Furthermore, as the Metaverse is user-centric by design, we need to always put the users' Quality of Service (QoS) as the core metric for these 360$^{\circ}$ video services. However, the QoS in 360$^{\circ}$ videos is much more complicated than traditional 2-Dimensional (2D) videos, and it is influenced by many factors~\cite{360videosurvey}. 360$^{\circ}$ videos can be viewed in either \textbf{\textit{VR}}~or~\textbf{\textit{non-VR}} mode. VR mode involves using a head-mounted display (HMD) to create an immersive experience, while non-VR mode refers to watching the video on a monitor~\cite{vrnonvr}. These two modes can result in very different quality of service (QoS) for users~\cite{vrnonvr}. With the emergence of the Metaverse, the boundary between VR and non-VR modes is becoming more blurred. While some users may prefer a fully immersive experience with an HMD, many others may opt for viewing 360$^{\circ}$ videos on their mobile or PC monitors at the same time. And due to the large data sizes of 360$^{\circ}$ video applications, it is necessary to utilize a server to generate and transmit video frames to users' display devices. Therefore, how to allocate resources to such different mixed users is a huge challenge.

\textbf{Challenges and motivations.} As the Metaverse aims to connect diverse users, QoS modeling for 360-degree videos is particularly challenging due to the heterogeneous users using both non-VR and VR modes. In VR mode, cybersickness is a critical issue that affects users' overall experience related to frame latency and stability~\cite{sicknesssurvey}. Additionally, perceptual video quality differs between different modes, even when having the same FPS, as indicated by studies~\cite{subjectivedata}. Designing a rational QoS model that accounts for the heterogeneous users in the Metaverse remains a significant challenge. Furthermore, most of the existing works focus on single-step optimization and use the optimized setting for the entire video, which is insufficient for efficient resource utilization. Our approach involves conducting a frame-wise optimization using a frame-slotted structure for the 360-degree video to decrease the latency fluctuations between frames and alleviate cybersickness. However, the frame-slotted sequential problem is non-convex and cannot be solved through separate single-step optimizations. We propose using Deep Reinforcement Learning (DRL) to address this time-sequential problem. However, traditional DRL methods may not provide a global allocation for the entire system in this heterogeneous scenario with two types of users. Therefore, there is an urgent need to develop a novel DRL method that considers users' preferences for resolution, latency, and frame rate to optimize the QoS.

\textbf{Related work and our novelty.} The QoS and QoE (Quality of Experience) in 360$^{\circ}$ videos contain many objective factors, such as the resolutions, frame rates, and frame delays~\cite{360videosurvey}. Researchers put much effort to designing the QoS model and optimize the objective metrics~\cite{360videosurvey}, e.g., Chen \textit{et al.} studied the QoS of a VR service over wireless communication using an echo state network~\cite{chenVR}. However, few of them has ever considered the QoS for the heterogeneous 360$^{\circ}$ video users (non-VR and VR users), or the cybersickness optimization for VR users. Besides, most of the works focus on single-step optimization and use this optimized setting for the whole video. On the contrary, we design the comprehensive QoS model for the mixed 360$^{\circ}$ video users, and a frame-slotted structure for optimizations in each frame, improving both kinds of users experience at the same time. Machine-learning-based approaches have been widely adopted to tackle wireless communication challenges~\cite{drlwcsurvey}, and DRL has been proven to achieve excellent performance. Nevertheless, few of them design a novel DRL approach with more specific view on different type of users. 

\textbf{Contributions.} Our contributions are as follows:
\begin{itemize}
    \item We design a frame-slotted structure and conduct frame-wise optimization, fully utilize the network resources.
    \item We craft a rational QoS model for heterogeneous 360$^{\circ}$ video users (non-VR and VR modes), involving the frame rate (FPS) and cybersickness optimization.
    \item We create two differentiated DRL structure and conduct comprehensive experiments. Our results demonstrate the superior performance of both methods in most heterogeneous scenario, achieving 15.2\%, 20.8\% improvement in frame rates for two types of users, and -80.9\% decrease in cybersickness for users in VR mode, compared to the traditional DRL algorithm.
\end{itemize}
\subsection{Organization} 
The remainder of the paper is structured as follows. In Section~\ref{sec:models}, we present our system model. Next, in Sections~\ref{sec:environment} and \ref{sec:algorithm}, we describe our deep reinforcement learning environment and proposed algorithms. Then, in Section~\ref{sec:experiments}, we conduct comprehensive experiments and compare our approach to various methods to demonstrate its effectiveness. Finally, in Section~\ref{sec:conclude}, we provide concluding remarks.

\section{System Model}
\vspace{-0.1cm}
\label{sec:models}

\begin{figure}[t] 
\centering
\setlength{\abovecaptionskip}{-0cm}
\includegraphics[width=1\linewidth]{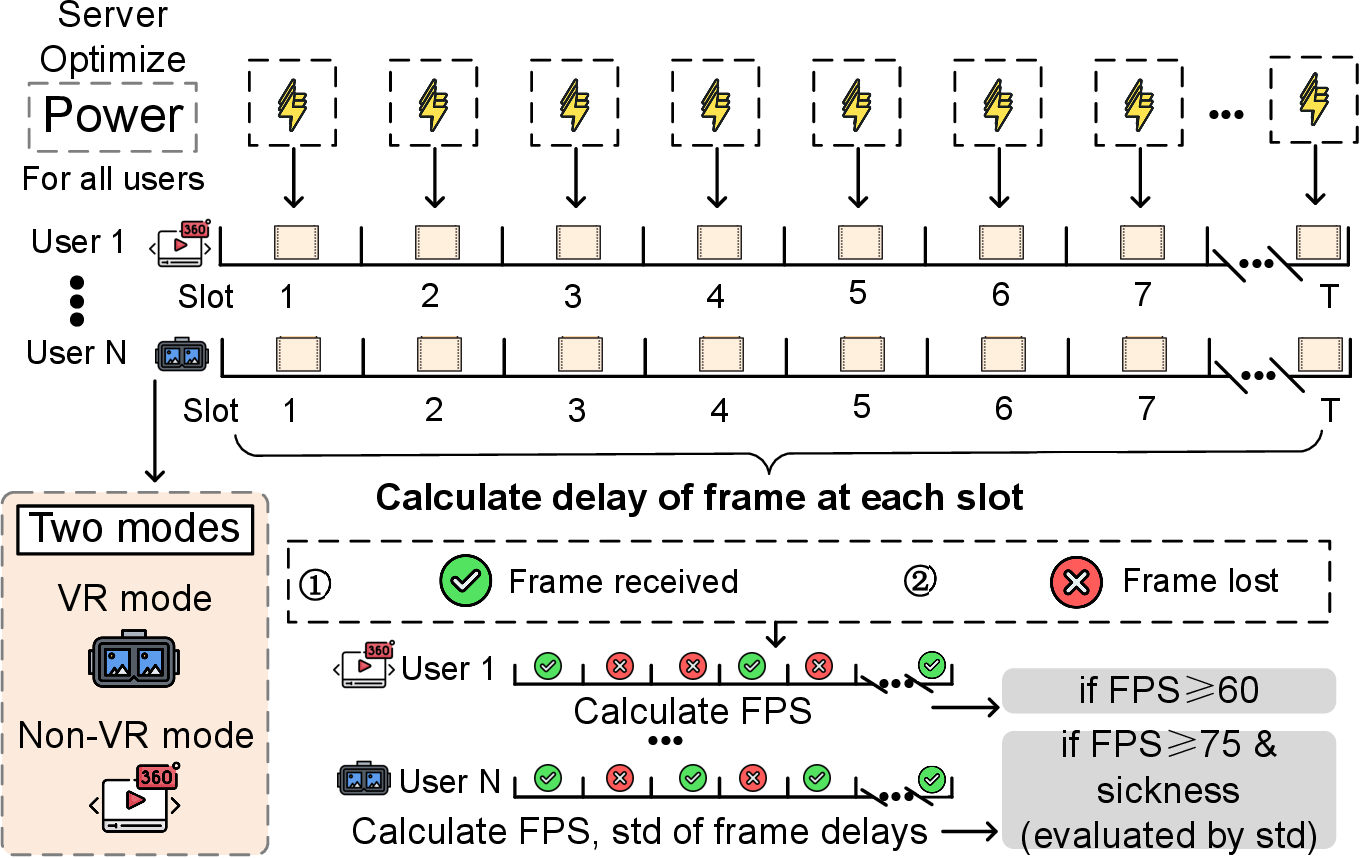}\vspace{-0.2cm}
\caption{Frame slotted transmission system.}\vspace{-0.6cm}
\label{fig:framedelay}
\end{figure}

\subsection{Frame-slotted structure}
This paper examines the downlink wireless transmission of 360-degree videos from a Video Server (VS) with multiple resolutions to a set of VR Users (VUs), represented by $\boldsymbol{N}$, who may be using different display types (HMD or monitor). $\boldsymbol{N}$ is defined as $\{1,2,\ldots,N\}$. Since the video transmission is typically transmitting a series of frames~\cite{videotransmission}, we employ a frame-slotted structure to ensure a seamless video experience. In this structure, the VS generates $\boldsymbol{T}$ frames ($\boldsymbol{T}=\{1,2,\ldots,T\}$) per second, and each second is partitioned equally into $T$ slots, with each slot transmitting one frame to the video users (VUs). This design helps to optimize the delivery of video content by ensuring that each frame is transmitted efficiently and received by the VUs in a timely manner. In each slot, the same frames will be transmitted from the VS to all VUs, and the Transmission Time Interval (TTI) of each slot it $\frac{1}{T}$ (one second with $T$ frames). To ensure the smooth experience for all VUs, we assume that the frame will be lost if it is not finished in its TTI.

\subsection{Propagation process}
At the beginning of each TTI $t$, the server will allocate the downlink transmission power for each VU. We use a $N\times T$ matrix $\boldsymbol{P}$ to denote the downlink transmission power allocated to VUs, where the element $p_n^t$ in $n^{\text{th}}$ row and $t^{\text{th}}$ column refers to the allocated power to VU $n$ ($n\in\boldsymbol{N}$) at slot $t$ ($t\in\boldsymbol{T}$), then, the server will transmit the frames to each VU. To alleviate the interference and simplify the transmission, we leverage frequency division multiple access (FDMA) in the propagation model. Therefore, the achievable transmission rate for VU $n$ at $t$ is:
\begin{align}
    r_n^t = W_n \log_2(1+\frac{p_n^tg_n^t}{\sigma^2W_n}), \label{shannon}
\end{align}
where $W_n$ denotes the bandwidth for VU $n$, $g_n^t$ is the channel gain of VU $n$ at TTI $t$, and $\sigma^2$ is the power spectral density of additive white Gaussian noise. Since the TTI is very short, we assume that $g_n^t$ remains constant within a given TTI, but varies across different TTIs. Then, the transmission latency $l_n^t$ for VU $n$ at TTI $t$ should be:
\begin{align}
    l_n^t = min\left(\frac{f\times b}{c_n^t\times r_n^t}, \frac{1}{T}\right), \label{eq:delay}  
\end{align}
where $f$ is the resolution (i.e., number of pixels), $b$ is the bits per pixel (bpp), $c_n^t$ is the compression ratio of this frame. Noted that compression ratios vary depending on the quality of the image and the size of the data, and it's always not constant~\cite{compression}. The use of $min()$ in this context implies that if the transmission delay is longer than the TTI, the frame will be lost and the latency will be capped at the value of TTI (i.e., $\frac{1}{T}$). We define a transmission success indicator as
    \begin{align}
        I_n^t = 
        \begin{cases}
            1, &\text{if}~~~\frac{f_n^t}{c_n^t\times r_n^t} \leq \frac{1}{T}. \\
            0, &\text{if}~~~\frac{f_n^t}{c_n^t\times r_n^t} > \frac{1}{T}.
        \end{cases}
    \end{align}
Then, the achievable frame rate of VU $n$ is $\sum\limits_{t=1}^TI_n^t$.

\subsection{QoS for VR and non-VR modes}
Each VU $n$ can use either VR mode (HMD) or non-VR mode (monitor), which is represented by a binary variable $\kappa_n \in \{0,1\}$, where $0$ denotes non-VR and $1$ denotes VR. We assume that the chosen mode remains constant during the whole process, because our optimization is conducted on a per-frame basis, with a very short time interval. The QoS varies significantly between different modes, particularly in terms of frame rate (measured in frames per second) and cybersickness~\cite{vrnonvr}. Thus, we explain our QoS model from these aspects according to the two modes.

\textit{Frame rate}: The ideal frame rates for VR and non-VR are different, because the HMD VR requires a higher frame rate to maintain a sense of presence and prevent motion sickness~\cite{sicknesssurvey}. Thus, we take the frame rate as a time-sequential optimization object in this paper, which is to make more frames be transmitted successfully in TTIs. And we set the minimum acceptable frame rates for VR and non-VR as $\Bar{I}_{\text{vr}}$ and $\Bar{I}_{\text{non}}$.

\textit{Cybersickness} (for HMD VR): Cybersickness is a complex phenomenon that is challenging to model accurately. One critical factor that affects cybersickness when using a \textbf{HMD VR} is the stability and latency between frames. It may seem counterintuitive, but the length of delays between each frame is not necessarily the most crucial factor. Instead, it is the abrupt high latency and delays fluctuations that can cause severe sickness for users~\cite{sicknesssurvey}. Therefore, simply minimizing the delays between frames may not be a feasible approach to reduce user discomfort, as it can potentially compromise resolution quality in order to achieve this trade-off. Our objective is to minimize the standard deviation (std) of delays between each pair of successfully transmitted frames. However, since cybersickness is not a concern in non-VR mode~\cite{vrnonvr}, we only need to consider this issue for VR mode.

To simplify the formulation of the delay standard deviation, we order the successfully transmitted frame (i.e., $I_n^t=1$) as $\{\mu_n^1, \mu_n^2, \ldots, \mu_n^K\}$ ($\forall n\in\boldsymbol{N}$), where $K=\sum\limits_{t=1}^T I_n^t$ means there are K frames successfully transmitted in total. $\mu_n^i=j$ ($i\in\{1, 2, \ldots, K\}$, $j\in\boldsymbol{T}$) means the $i^{\text{th}}$ successful frame is the $j^{\text{th}}$ frame among all frames in one second. To provide a clearer explanation of the delays between each pair of received frames, we have included a visual aid in Fig.~\ref{fig:framedelay}. Once the $i^{\text{th}}$ frame is successfully received at slot $\mu_n^i$, we determine the time interval until the next successful frame is received. Then, the delay between every two successful frames $i$ and $i+1$ is:
\begin{align}
    d_n^{i,i+1} = (\frac{1}{T}-l_n^{\mu_n^i})+\sum_{t=\mu_n^i+1}^{\mu_n^{i+1}}l_n^t, \forall n\in\boldsymbol{N}.
\end{align}
And the standard deviation of delays between each pair of frames for VU $n$ is:
\begin{align}
    \text{std}_n = \sqrt{\frac{\sum_{i=1}^{K-1} (d_n^{i,i+1}-\Bar{d}_n)^2}{K-1}},
\end{align}
where $\Bar{d}_n$ is the mean of these delays $\frac{\sum_{i=1}^{K-1} d_n^{i,i+1}}{K-1}$.

\begin{figure}[t] 
\centering
\setlength{\abovecaptionskip}{-0.1cm}
\includegraphics[width=1\linewidth]{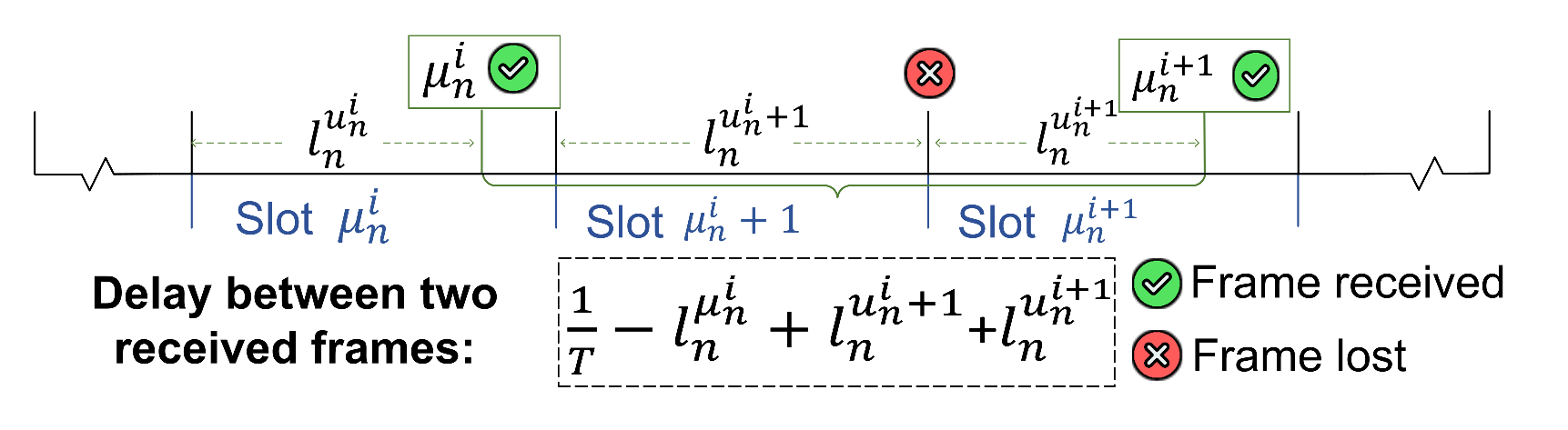}
\caption{The delay between two successful frames.}\vspace{-0.4cm}
\label{fig:framedelay}
\end{figure}

Our QoS model integrates both frame rate and cybersickness, which is as follows:
\begin{align}
    Q_n = \omega_1 \underbrace{\sum_{t=1}^T I_n^t}_{\text{frame rate}} - 
    \omega_2 \underbrace{\kappa_n \text{std}_n}_{\text{cybersickness}},
\end{align}
where $\omega_1, \omega_2$ are weight parameters for these three metrics, which will be numerically defined in Section~\ref{sec:environment}. Note that $\kappa_n$ is multiplied in the delay std part as cybersickness is only considered when the user is using VR mode.

\subsection{Problem formulation}
This paper aims to optimize transmission power $\boldsymbol{P}$ during $T$ frames, to maximize the QoS for all VUs, subject to the frame rate and power constraints. From above discussion, the formulated problem is:
\begin{align}
    &\max\limits_{\boldsymbol{P}} \sum_{n\in\boldsymbol{N}} Q_n.\\
    &s.t.~C1: \sum_{t=1}^T I_n^t \geq (1-\kappa_n) \Bar{I}_{\text{non}}, \forall n\in \boldsymbol{N},\\
    &~~~~C2: \sum_{t=1}^T I_n^t \geq \kappa_n \Bar{I}_{\text{vr}}, \forall n\in \boldsymbol{N},\\
    &~~~~C3: \sum_{n\in\boldsymbol{N}} p_n^t \leq p_{max}, \forall t\in\boldsymbol{T}.
\end{align}
Constraint $C1,C2$ are the frame rate requirements for non-VR and VR modes, respectively. These constraints will be fulfilled by applying an early-termination flag to the DRL training, which will be explained in detail in Section~\ref{sec:experiments}. Constraint $C3$ is the sum-power limit.

\textbf{Why DRL method?} The formulated problem is non-convex and NP-hard. Its time-sequential nature involving cybersickness and frame rate optimization across multiple time slots significantly increases the number of variables with respect to time, denoted as $T$. Therefore, using the traditional convex optimization strategy is not appropriate in such a complicated situation. Besides, heuristic search, while naively improving action selection based on the current value function, it typically involves considering a vast tree of potential continuations~\cite{RLintro}, which makes it impractical to apply heuristic search in scenarios with a large number of decisions and dimensions to consider. The effectiveness in handling time sequential problems of the DRL method has been demonstrated in numerous studies~\cite{drlwcsurvey}. With appropriate reward settings and algorithm structure design, this method can achieve satisfactory results in complex scenarios.

\section{DRL Environment Desgin} \label{sec:environment}
When solving problems with DRL methods, the foremost step is to craft a comprehensive DRL environment for the DRL agent, including three key elements: state, action, and reward.

\textbf{State.} The global state $s_g^t=\{s_{\text{non}}^t; s_{\text{vr}}^t\}$ contains two sets of users' states: the non-VR users' states $s_{\text{non}}^t$ and VR users' states $s_{\text{vr}}^t$. For non-VR user states $s_{\text{non}}^t$, it involves 
\begin{itemize}
    \item left frames needed to be transmitted $T-t$
    \item left tolerant failure times $(T-\Bar{I}_{\text{non}})-\sum_{t'=1}^t (1-I_n^t$)
    \item the data size of current frame $D_n^t = \frac{f\times b}{c_n^t}$
    \item the current channel gain $g_n^t$
\end{itemize}
($\forall n~\text{that}~\kappa_n=0$). In addition to the elements mentioned above, the states $s_{\text{vr}}^t$ for VR users also include another crucial factor: $\text{std}_n^t$, which represents the standard deviation of current delays between successful frames.

\textbf{Action.} The action $a^t$ in this paper is the allocated transmission powers for each VU $n$ used to communicate with the server: $a^t = \{p_1^t, p_2^t, \ldots, p_N^t\}$. However, it is not feasible for the RL algorithm to allocate power for all VUs within the summed power constraint. Therefore, we incorporate a softmax layer into the network to convert the output into fractions $\{{p'}_1^t, {p'}_2^t, \ldots, {p'}_N^t\}$ (with a sum of 1). Afterwards, we multiply these fractions by $p_{max}$ (the maximum power resource of the server).

\textbf{Rewards.} Rational reward setting is crucial for DRL training. Here, we give our numerical reward setting for reference. Similar to the state, the reward is also divided into $R_{\text{non}}^t$ for non-VR users and $R_{\text{vr}}^t$ for VR users. The $R_{\text{non}}^t$ includes: (1) frame success reward $R_{f,non}^t$: $+1$ for all VUs on every success. (2) the early termination penalty (the left tolerant failure time is $0$): $-2\times (T-t)$. While the $R_{\text{vr}}^t$ comprises: (1) frame success reward $R_{f,vr}^t$: $+1.5$ for each as the VR mode is more sensitive to frame loss. (2) early termination penalty $-2\times (T-t)$. (3) the frame delay std penalty $R_{std}^t$: $-1000\times\text{std}_n^t$ (order of magnitude is generally $10^{-3}$).

\section{Methodology} \label{sec:algorithm}
Our algorithm is based on Proximal Policy Optimization (PPO)~\cite{PPO}, which serves as the backbone. We then introduce novel structures that are tailored to our specific problem. In this section, we provide a brief overview of PPO and then elaborate on our customized algorithms that build upon it.

\subsection{Backbone-PPO}
The Actor-Critic-based algorithm is currently regarded as the most advanced and effective DRL algorithm, employing an Actor network to choose actions and a Critic network to evaluate them. One state-of-the-art algorithm within this framework is Proximal Policy Optimization (PPO), which has demonstrated impressive performance across a range of scenarios, including the widely discussed ChatGPT~\cite{ChatGPT}. PPO incorporates two key techniques in its policy network (Actor): (1) Importance Sampling, which is using the current policy for sampling trajectories, and the previous policy for calculating the action advantages (how is the current state), to increase the sample efficiency. (2) KL penalty between the current and previous policies to enhance stability. Therefore, we use PPO as the backbone of our proposed algorithms. For the sake of brevity, we will only provide the update functions of the Actor and Critic networks. A detailed, step-by-step explanation of PPO can be found in our previous work~\cite{AAHC}.

The Actor is updated by gradient ascent, and the update function is~\cite{PPO}:
\begin{align}
    \Delta\theta = \mathbb{E}_{(s^t,a^t)\sim\pi_{\theta_{'}}}[\triangledown f^t(\theta,A^t)], \label{eq:actorobj}
\end{align}
where $f^t(\theta)=min\{r^t(\theta)A^t, clip(r^t(\theta), 1-\epsilon, 1+\epsilon)A^t\}$ serves as the policy change constraint. $\pi_{\theta}, \pi_{\theta'}$ are the current and previous policies, and the $r^t(\theta)=\frac{\pi_\theta(a^t|s^t)}{\pi_{{\theta'}}(a^t|s^t)}$ is the ratio between the two policies. $\epsilon$ is the clip rate. 

The $A^t(R^t, s^t)$ (short as $A^t$) denotes the advantage of state that is calculated by rewards and states. We use the state-of-art generalized advantage estimation (GAE)~\cite{schulman2015high} to calculate the advantage function:
\begin{align}
    &A^t = \delta^t + (\gamma\lambda)\delta^{t+1}+...+(\gamma\lambda)^{\bar{T}-1}\delta^{t+\bar{T}-1}, \\
    &\text{where}~~~\delta^t=R^t+\gamma V_{\phi'}(s^{t+1})-V_{\phi'}(s^t).
\end{align}

In terms of the value network (Critic), PPO uses identical Critic as per other Actor-Critic algorithms; and the loss function can be formulated in~\cite{PPO} as:
\begin{align}
    L(\phi) = [V_\phi(s^t)-(A^t+V_{\phi'}(s^{t}))]^2. \label{eq:criticloss}
\end{align}

The state-value function $V(s)$, as described in~\cite{RLintro}, is a commonly used metric that is estimated by a learned Critic network with parameter $\phi$. To update $\phi$, we minimize $L(\phi)$ and periodically update the parameter $\phi'$ of the target state-value function with $\phi$, a technique known as target value, which is prevalent in RL~\cite{RLintro}.

\subsection{Our proposed methods}
Despite the advantages of PPO, its traditional structure may not be sufficient for our formulated problem. Our scenario involves a highly mixed population of non-VR and VR mode users, which presents a heterogeneous situation for the RL agent since it must output actions (i.e., downlink transmission powers) for all VUs simultaneously. Using the standard PPO structure in this situation can result in a heterogeneous and blurred total reward as feedback for the agent, similar to the sparse-reward problem~\cite{sparsereward}, which can significantly impede DRL training. Consequently, we have developed two unique structures based on PPO to address our problem. 

In order to evaluate the current state for non-VR and VR users separately and obtain the corresponding values $V_{\text{non}}^t$ and $V_{\text{vr}}^t$, we employ a differentiated structure to leverage domain knowledge and accelerate training speed. Accordingly, we modify the update functions of Actor in Eq.~(\ref{eq:actorobj}) and Critic in Eq~(\ref{eq:criticloss}):
\begin{align}
    &\text{Actor: }\Delta\theta = \mathbb{E}_{(s^t,a^t)\sim\pi_{\theta_{'}}}[\triangledown f^t(\theta, (A_{\text{non}}^t+A_{\text{vr}}^t))], \label{eq:actorobj2}\\
    &\text{Critic: }L^t(\phi) = [V_{\text{non}}^t-(A_{\text{non}}^t+{V'}_{\text{non}}^t)]^2 +\nonumber\\
    &~~~~~~~~~~~~~~~~~~[V_{\text{vr}}^t-(A_{\text{vr}}^t+{V'}_{\text{vr}}^t)]^2
\end{align}
where
\begin{align}
    &A_{\text{non}}^t=\delta_{\text{non}}^t + (\gamma\lambda)\delta_{\text{non}}^{t+1}+...+(\gamma\lambda)^{\bar{T}-1}\delta_{\text{non}}^{t+\bar{T}-1},\\
    &\delta_{\text{non}}^t=R_{\text{non}}^t+\gamma {V'}_{\text{non}}^{t+1}-{V'}_{\text{non}}^t,\label{eq:delta1}\\
    & \text{and the same for } A_{\text{vr}}^t. \nonumber
\end{align}

The updated Actor function Eq.(\ref{eq:actorobj2}) sums the advantages $A_{\text{non}}^t$ and $A_{\text{vr}}^t$, which are calculated by the Critic for non-VR and VR users, respectively. The updated Critic function Eq.(\ref{eq:criticloss}) evaluates the values $V_{\text{non}}^t$ and $V_{\text{vr}}^t$ (value with prime is evaluated by the target network), and updates by summing the losses from both values. The approach of summing the losses is inspired by the Hybrid Reward Architecture~\cite{HRA}.

\textbf{How to get the values?} The process for the Actor is straightforward: it takes in the global state as input and generates actions for all users. However, the way the Critic evaluates the two values for non-VR and VR users is quite different. In other words, it is unclear how to obtain $V_{\text{non}}^t$ and $V_{\text{vr}}^t$. To address this issue, we propose two different approaches that can be used to obtain these values.

\noindent \textbf{Approach 1: Separate Input Differentiated Output (SIDO)}
We provide two separate states, $s_{\text{non}}^t$ and $s_{\text{vr}}^t$, to the Critic, which can be thought of as two Critic branches that share the same upper layers. The first branch takes $s_{\text{non}}^t$ as input and outputs the value for non-VR users, while the second branch takes $s_{\text{vr}}^t$ as input and outputs the value for VR users:
\begin{align}
V_{\phi}^{\text{non}}(s_{\text{non}}^t)=V_{\text{non}}^t; V_{\phi}^{\text{vr}}(s_{\text{vr}}^t)=V_{\text{vr}}^t,
\end{align}
where $V_{\phi}^{\text{non}}$ and $V_{\phi}^{\text{vr}}$ are the two Critic branches.

\noindent \textbf{Approach 2: Merged~ Input Differentiated Output (MIDO)}
In this approach, we give the Critic the global state $s_g^t={s_{\text{non}}^t;s_{\text{vr}}^t}$, and the critic outputs $V_{\text{non}}^t, V_{\text{vr}}^t$ simultaneously, 
\begin{align}
    V_{\phi}(s_g^t)=\{V_{\text{non}}^t, V_{\text{vr}}^t\}.
\end{align}

Here, we use the term "differentiated" to refer to the Critic that can evaluate the two separate sets of users. The SIDO approach enables the Critic to focus on the specific aspects of the environment that are relevant to each user group (non-VR/VR), but it lacks global information as each branch of the Critic only has access to a subset of the input state. On the other hand, MIDO allows the Critic to consider all aspects of the environment, including those that are relevant to both user groups, but may struggle to learn specialized representations for each group. Thus, empirical testing will be necessary to determine which approach is more effective.

\begin{figure}[t] 
\centering
\setlength{\abovecaptionskip}{-0.1cm}
\includegraphics[width=1\linewidth]{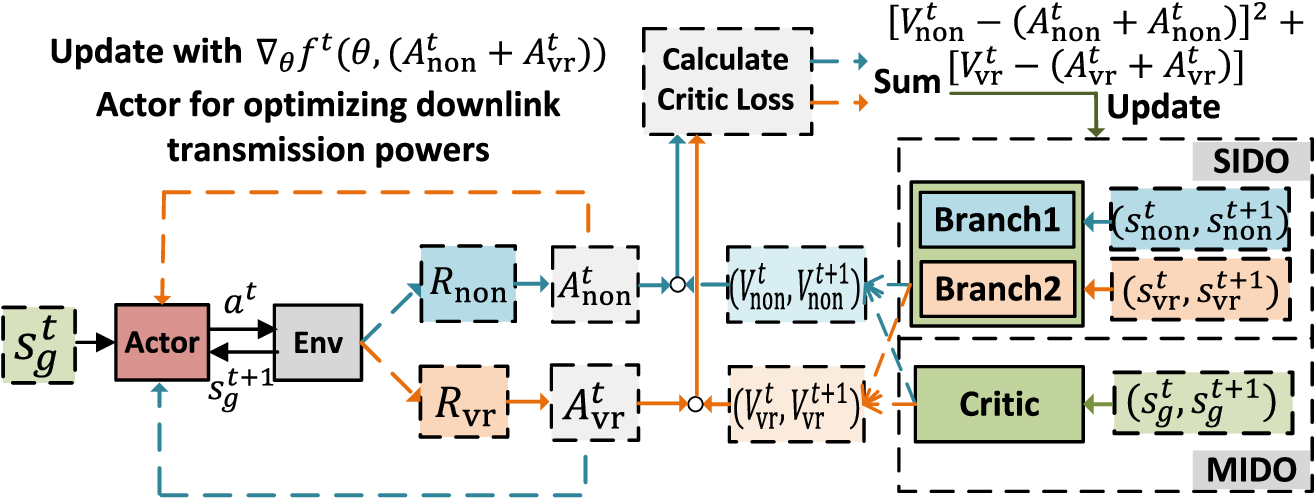}
\caption{Proposed SIDO and MIDO algotirhms.}\vspace{-0.6cm}
\label{fig:alg}
\end{figure}

\subsection{Baselines}
We will evaluate the performance of our proposed algorithms against the following baselines:
\begin{itemize}
    \item SIDO: proposed algorithm described above, which uses a separate input and differentiated output Critic PPO.
    \item MIDO: proposed algorithm described above, which uses a merged input and differentiated output Critic PPO.
    \item PPO: standard PPO algorithm with a single input (global state) and a single output (value for all users).
    \item Average allocation: a naive baseline that allocates downlink transmission power equally across all users.
\end{itemize}

In terms of the metrics, we select (1) frame rate, (2) delay std among VR users, and (3) successful steps (before left tolerance frame failure times run out) during training.

\textbf{Computational complexity:} 
We use $m^l_{A}, m^l_{C}$ to denote the number of neurons in layer $l$ of the Actor and the Critic. And $d(s)$ as the input layer (proportional to the state dimension), $(L_A, L_C)$ is the number of training layers of the three parts. Considering the mini-batch size $B$ in the training stage, we have the complexity in one training step as
\begin{wrapfigure}{r}{0.24\textwidth}
\vspace{-0.25cm}\includegraphics[width=0.24\textwidth]{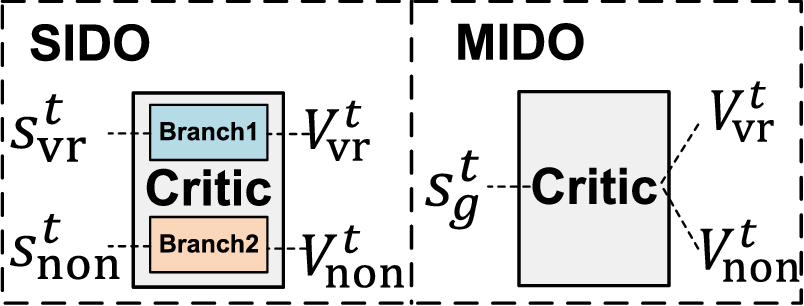}
\vspace{-430pt}
\end{wrapfigure}
\begin{align}
    &O(B( \underbrace{d(s_g^t)m^1_{A}\hspace{-3pt}+\hspace{-6pt}\sum_{l=1}^{L_{A}-1}\hspace{-3pt}m^l_{A}m^{l+1}_{A}}_{\text{Actor}}\nonumber\\
    &+\underbrace{\begin{cases}                   [d(s_{\text{vr}}^t)+d(s_{\text{non}}^t)]m^1_{C}+\sum_{l=1}^{L_{C}-1}m^l_{C}m^{l+1}_{C})), &\text{SIDO},\\
    d(s_g^t)m^1_{C}+\sum_{l=1}^{L_{C}-1}m^l_{C}m^{l+1}_{C})), &\text{MIDO}.
    \end{cases}}_{\text{Critic}} \label{eq:complexity}
\end{align}
And according to~\cite{complexity}, the total computational complexity depends on the total number of convergence steps to the optimal policy.
\label{sec:RLenv}

\section{Simulation} \label{sec:experiments}
In this section, we first briefly explain the numerical settings, and then evaluate the proposed algorithms.
\subsection{Numerical Settings}
We use 8 VUs in total with VR users from 2-6. Frame resolution is 2k with 16 bpp, and the compression ratio is uniformly selected from 300-500. The total frames per second are 90, and the bandwidth per channel is $10^6$ Hz. The required successful frame rates are $75$ for VR users and $60$ for non-VR users. Small-scale fading follows Rayleigh distribution and the path loss exponent is $2$. We train for $5\times10^5$ steps with evaluation every 50 steps. Experiments are conducted with the same global random seeds from 0-10 and error bands are included.

\subsection{Result Analysis}

\begin{figure}[t]
\centering
\subfigtopskip=2pt
\subfigbottomskip=2pt

\subfigure[Reward in 4 non-VR$|$VR users scenario during training.]{
\begin{minipage}[t]{0.48\linewidth}
\centering
\includegraphics[width=1\linewidth]{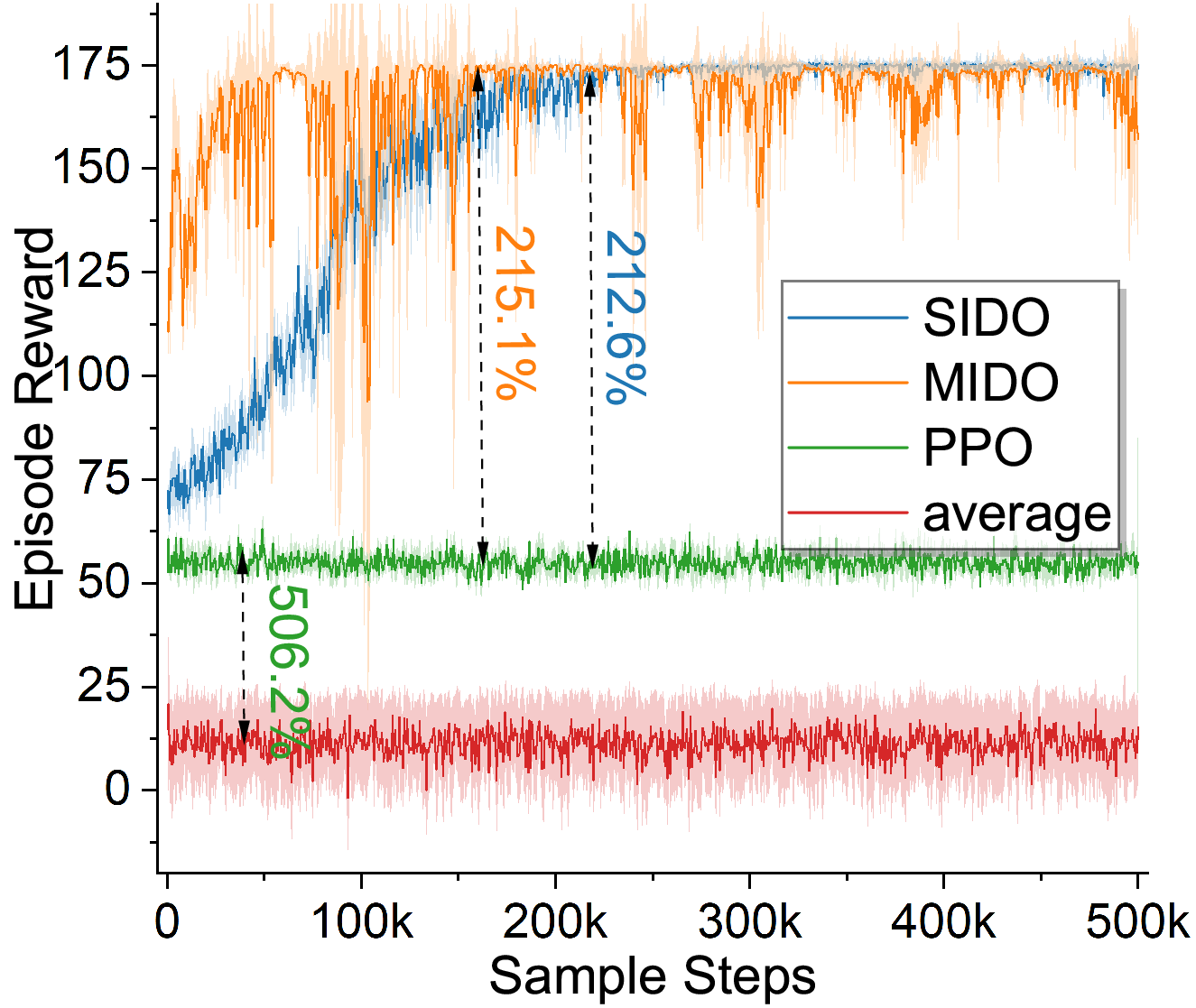}
\label{fig:reward}
\vspace{-10mm}
\end{minipage}
}%
\subfigure[Train \& Execution Time in different scenarios.]{
\begin{minipage}[t]{0.48\linewidth}
\centering
\includegraphics[width=1\linewidth]{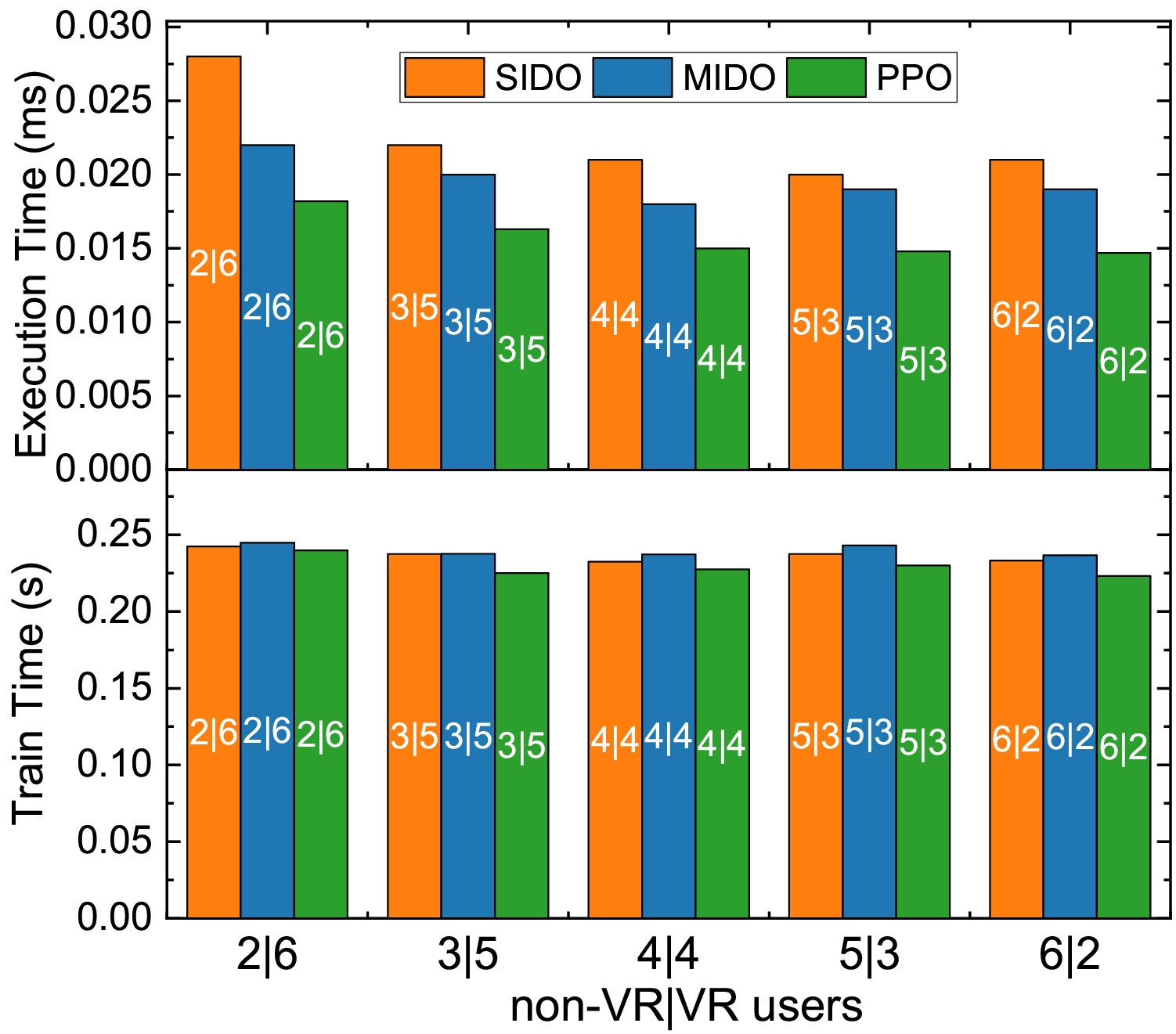}
\label{fig:time}
\vspace{-10mm}
\end{minipage}
}%
\vspace{-0.1cm}
\caption{Reward (left) and the train \& execution time (right). Considering the random evolving variables, all experiments were conducted using the same 10 global random seeds, and error bands were included in the illustrations.}
\label{fig:train}
\vspace{-0.5cm}
\end{figure}
\begin{figure*}[t]
\centering
\subfigtopskip=2pt
\subfigbottomskip=2pt
\subfigure[Average Frame Rate for non-VR (left) and VR (right) users.]{
\begin{minipage}[t]{0.61\linewidth}
\centering
\includegraphics[width=1\linewidth]{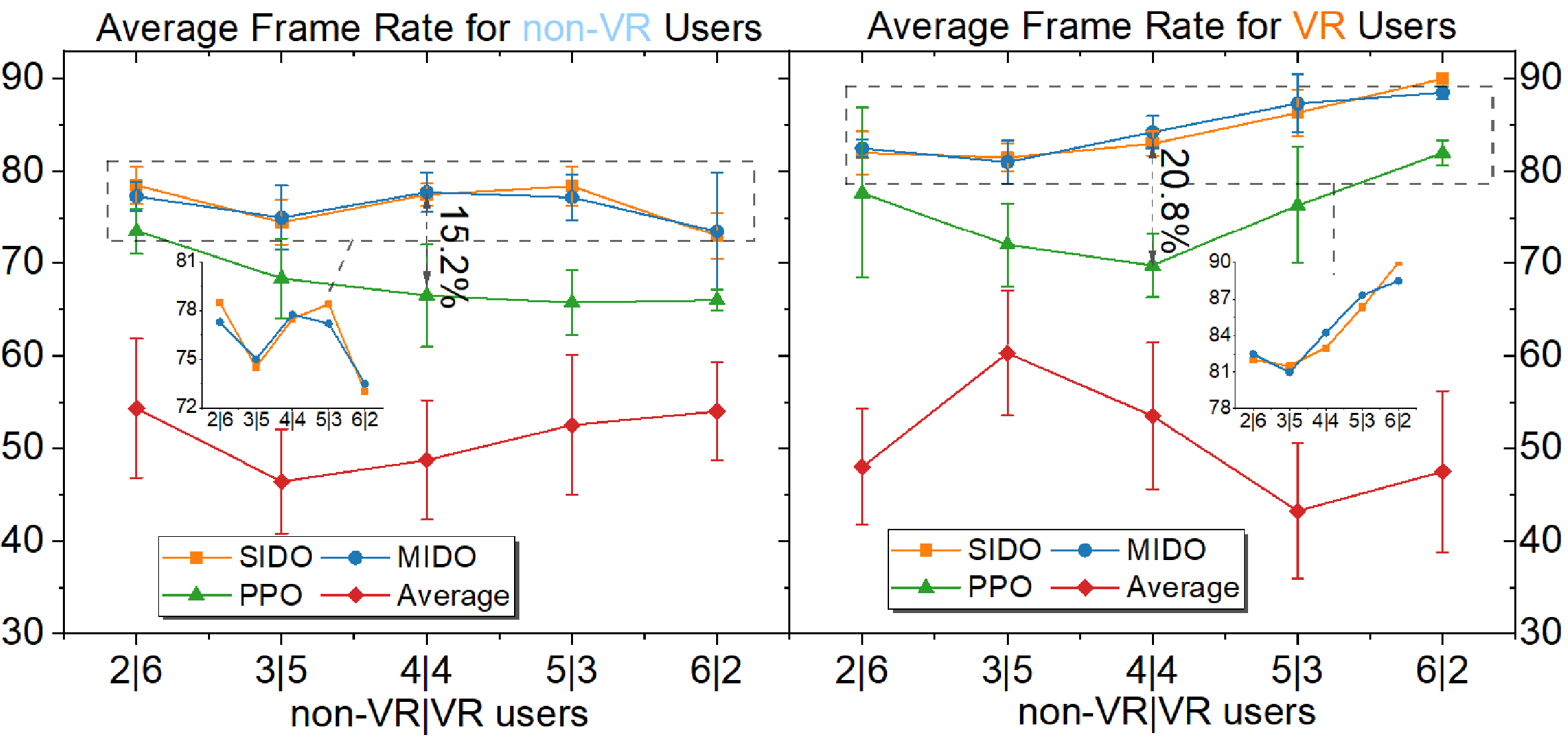}
\label{fig:fps}
\vspace{-10mm}
\end{minipage}
}%
\subfigure[Average Cybersickness for VR users.]{
\begin{minipage}[t]{0.3\linewidth}
\centering
\includegraphics[width=1\linewidth]{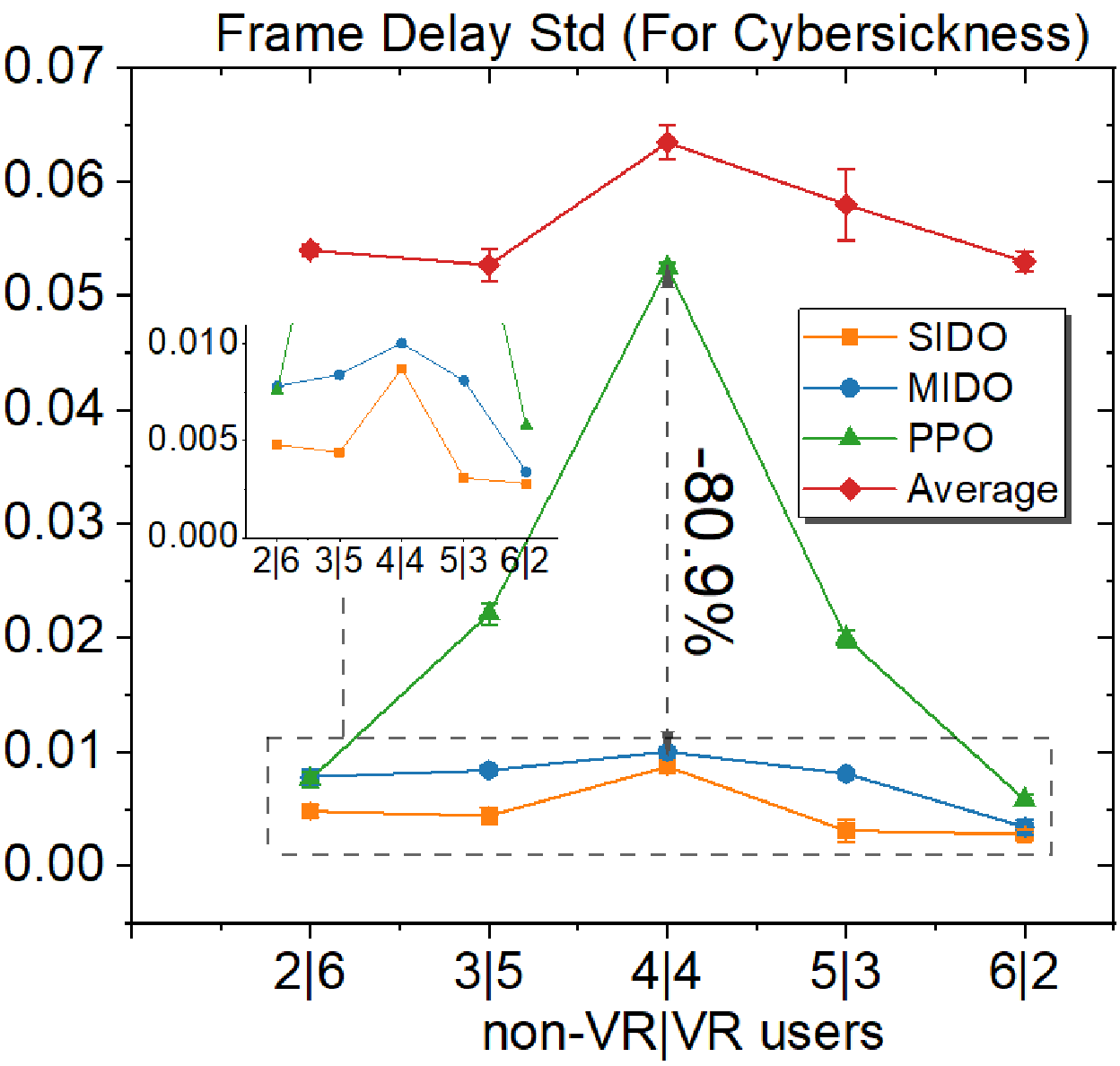}
\label{fig:sickness}
\vspace{-10mm}
\end{minipage}
}%

\caption{Metrics (i.e., achievable FPS for non-VR, FPS for VR users, and cybersickness for VR users) in different scenarios.}
\label{fig:train}
\vspace{-0.5cm}
\end{figure*}
Due to space limitations, we present the complete reward for a typical scenario of 4 non-VR users and 4 VR users (i.e., $4|4$) during training in Figure~\ref{fig:reward}. This scenario is representative of the challenges posed by heterogeneous user environments. In this scenario, both SIDO and MIDO outperform traditional PPO by a significant margin. Specifically, MIDO achieves approximately $215.1\%$ and SIDO obtains around $212.6\%$ improvement over PPO. In contrast, PPO drops into a local minimum in an early stage, similar to the common problem encountered when using the normal algorithm in other sparse-reward scenarios~\cite{sparsereward}. Although MIDO and SIDO reach a similar peak, there are also interesting differences between them during training, and these differences are reflected in other scenarios as well. It is evident that MIDO has a faster convergence speed (around $73,000$ sample steps) than SIDO, but SIDO is much more stable than MIDO. We assume that the fast convergence speed of MIDO can be attributed to its global view, while the stability of SIDO may benefit from the specific view of the states for different types of users. With the global view of all users, MIDO manages to find the optimal solutions, but the estimations of the values for two types of users may be not so accurate, since it only takes in the global state, containing mixed elements of both kinds of users, which causes instability. In contrast, SIDO is more stable because it takes into account the specific views of the states for different types of users. Furthermore, although PPO falls into the local minimum very soon, it still has a $506.2\%$ improvement over average allocation. We also illustrate the single step train \& execution time of different algorithms in Fig~\ref{fig:time}. Based on the computation complexity in Eq.~(\ref{eq:complexity}), the computation complexity of SIDO is only slightly greater than that of MIDO since $d(s_{\text{vr}}^t)+d(s_{\text{non}}^t)$ is greater than $d(s_g^t)$ due to the duplicated elements. Moreover, the state dimension is influenced by different scenarios, as the VR users state contains the cybersickness element. In general, the increment of both train \& execution time of MIDO and SIDO compared to PPO is acceptable.

Fig.~\ref{fig:fps} and Fig.~\ref{fig:sickness} illustrate the achievable FPS (frames per second) and cybersickness (frame delays std) for users. For the FPS, SIDO and MIDO have similar performance, obtaining $15.2\%$, $20.8\%$ improvements among non-VR users and VR users than PPO, respectively, in the $4|4$ scenario. This is the scenario with most heterogeneous users, and the gaps in other scenarios between SIDO, MIDO and PPO are smaller. The similar results are also sown for the cybersickness, while SIDO is slightly better than MIDO according to cybersickness. We think this is also due to the specific view of SIDO, since the cybersickness is only considered for VR users, and SIDO can get a more clear view on the VR users.

\section{Conclusion} \label{sec:conclude}
This paper create a frame-slotted structure and conduct the frame-wise optimization, considering the scenario with heterogeneous 360-degree video for both non-VR and VR users. We optimize the frame rates for both kinds of users, and cybersickness for VR users. In general, our proposed SIDO and MIDO both achieve much superior performance compared to traditional PPO algorithms over the achievable FPS and cybersickness. In the future, we will try to optimize the allocation of resolutions for each frame simultaneously, and extend the proposed SIDO, MIDO to multi-agent DRL structures.

\renewcommand{\refname}{~\\[-20pt]References\vspace{-5pt} }

\bibliographystyle{IEEEtran}

\end{document}